# Virtual H&E Histology by Fiber-Based Picosecond Two-Photon Microscopy

Jan Philip Kolb[1], Daniel Weng[2], Hubertus Hakert[1], Matthias Eibl[2], Wolfgang Draxinger[2], Tobias Meyer[3], Thomas Gottschall[4], Ralf Brinkmann[1], Reginald Bringruber[2], Jürgen Popp[3], Jens Limpert[4], Sebastian Nino Karpf[2], Robert Huber[2]

[1]Medizinisches Laserzentrum Lübeck GmbH, Peter-Monnik-Weg 4, 23562 Lübeck, Germany

[2]Institut für Biomedizinische Optik, Universität zu Lübeck, Peter-Monnik-Weg 4, 23562 Lübeck, Germany

[3]Institut für Physikalische Chemie, Friedrich Schiller Universität Jena, Helmholtzweg 4, 07743 Jena, Germany

[4]Institut für angewandte Physik, Friedrich Schiller Universität Jena, Albert-Einstein-Straße 15, 07745 Jena, Germany

**ABSTRACT**

Two-Photon Microscopy (TPM) can provide three-dimensional morphological and functional contrast *in vivo*. Through proper staining, TPM can be utilized to create virtual, H&E equivalent images and thus can improve throughput in histology-based applications. We previously reported on a new light source for TPM that employs a compact and robust fiber-amplified, directly modulated laser. This laser is pulse-to-pulse wavelength switchable between 1064 nm, 1122 nm, and 1186 nm with an adjustable pulse duration from 50ps to 5ns and arbitrary repetition rates up to 1MHz at kW-peak powers. Despite the longer pulse duration, it can achieve similar average signal levels compared to fs-setups by lowering the repetition rate to achieve similar cw and peak power levels. The longer pulses lead to a larger number of photons per pulse, which yields single shot fluorescence lifetime measurements (FLIM) by applying a fast 4 GSamples/s digitizer. In the previous setup, the wavelengths were limited to 1064 nm and longer. Here, we use four wave mixing in a non-linear photonic crystal fiber to expand the wavelength range down to 940 nm. This wavelength is highly suitable for imaging green fluorescent proteins in neurosciences and stains such as acridine orange (AO), eosin yellow (EY) and sulforhodamine 101 (SR101) used for histology applications. In a more compact setup, we also show virtual H&E histological imaging using a direct 1030 nm fiber MOPA.

## 1. INTRODUCTION

Two-photon excitation fluorescence microscopy (TPM) [1] is a powerful technique for sensitive tissue imaging with strong depth sectioning capabilities compared to standard microscopy. The unique combination of deep penetration in highly scattering media like brain or skin of up to 1000 micrometers and superior optical resolution has proven to be valuable in fields like neuronal imaging [2] and dermatology. More recently, it has also been shown that it can be used for creating virtual hematoxylin and eosin (H&E) like images of tissue without the time- and labour-intense process of cutting the tissue into µm-thin slices [4] among other methods such as light-sheet microscopy [5], stimulated Raman scattering microscopy [6] and microscopy with ultraviolet surface excitation (MUSE) [7].

Apart from the commonly used femtosecond laser systems, we recently presented TPM with nanosecond and picosecond pulses (nsTPM) at low repetition rates. We showed that equivalent images can be obtained with the same cw-power on the sample if the duty cycle is the same as for femtosecond lasers [8]. TPM with nanosecond pulses offers several advantages over standard TPM with femtosecond pulses: In contrast to femtosecond pulses, nanosecond pulses experience practically no pulse stretching due to dispersion, therefore, they can be easily guided through optical fibers in contrast to fs-pulses. This results in a robust and low-footprint setup that can be used in a clinical environment. We also demonstrated several other advantages of our method such as pulse to pulse wavelength switching [9] and single pulse fluorescence life time imaging (FLIM) measurements [10]. Additionally, our fiber master oscillator power amplifier (fiber MOPA) is ytterbium based and therefore can be easily scaled in cw power at low cost and it offers pulse on demand capabilities that could be used for advanced sampling protocols. We plan to apply nsTPM to virtual H&E and neuronal imaging

Virtual H&E imaging requires staining to create a contrast between the nuclei and the remaining structures. In conventional H&E histology the nuclei are stained with hematoxylin and the counterstain eosin. Eosin itself is fluorescent and can therefore be used for TPM, but hematoxylin is not fluorescent and hence needs a replacement. We decided to use the staining protocol suggested by Cahill et al. [11] with acridine orange (AO) as a nuclear stain and sulforhodamine 101 (SR101) as a counterstain. These two fluorophores are excitable with ~1030 nm or shorter wavelengths, but not with 1064 nm as shown below. Therefore, we had to modify our fiber MOPA to emit at 1030 nm or shorter.

Additionally, we intend to demonstrate neuronal imaging with our system. The widely-used green fluorescent protein (GFP) [12] for investigating neuronal signaling requires an excitation at ~940 nm. This new wavelength can be generated by degenerate four-wave mixing (FWM) in a large-mode-area photonic-crystal fiber (PCF) [13]. We are going to apply this to our fiber MOPA laser to obtain wavelengths suitable for GFP excitation.

In this publication, we demonstrate a fiber MOPA with 940nm wavelength generated via FWM in a PCF. Furthermore, we present a fiber MOPA with 1030 nm emission wavelength. Virtual H&E imaging with this laser is performed and the results are compared to images created with a commercial two photon microscope using a femtosecond Titanium Sapphire (Ti:Sa) laser. We investigate the choice of the excitation wavelength and the choice of emission filters for AO and SR101.

## 2. METHODS

In this section, we describe the setup used for the generation of 940 nm pulses by FWM in a PCF, our home-built picosecond multiphoton microscopy setup including its fiber master oscillator power amplifier (fiber MOPA), the commercial femtosecond multiphoton microscopy setup used for comparison, the staining procedure of the tissue and the digital processing.

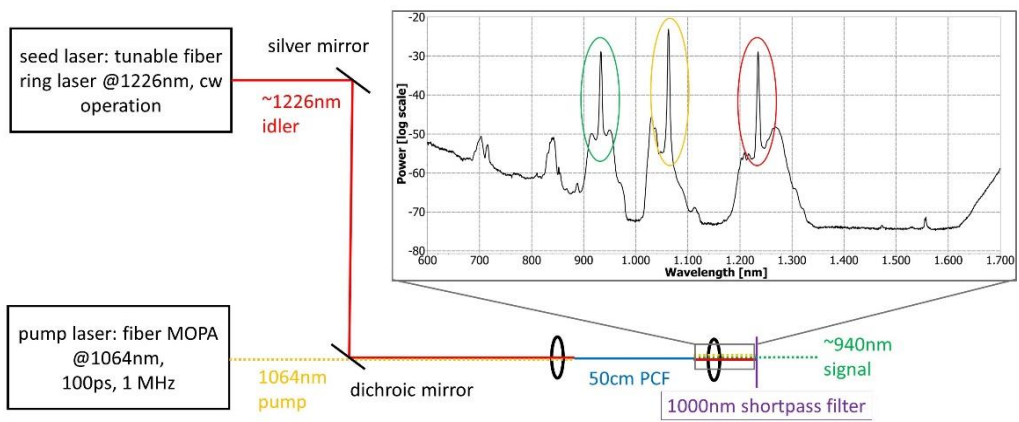

Figure 1: Setup for 940nm light generation. By coupling both a 1064 nm pump and a 1226 nm idler light into a Photonic Crystal Fiber (PCF), phase-matching leads to high-power pulses being generated at 940 nm. The resulting spectrum directly after the PCF is shown in the top right corner visualizing the three wavelengths contributions at 940 nm (green), 1064 nm (yellow) and 1226 nm (red).

The setup used for generation of 940 nm via FWM is displayed in Figure 1 and is adapted from [13]. It requires a pump laser and a seed laser, which acts as the idler in FWM process. The pump laser is a fiber MOPA emitting 100 ps pulses at 1 MHz repetition rate and 1064 nm wavelength. It was previously presented by our group in [14]. The seed laser is a tunable fiber ring laser consisting of a semiconductor optical amplifier as a gain medium (Innolume SOA-1250-110-YY-27dB), a Fabry-Pérot filter, an optical isolator and a fiber coupler. Both lasers are combined by a dichroic mirror and coupled into ~50 cm of PCF (NKT Photonics LMA-PM-5). This PCF provides a strong non-linearity and phase-matching condition to efficiently generate 940nm light via FWM. The exact output wavelength at 940nm can be tuned by tuning the seed laser around 1226nm. This idler seed laser can be operated in continuous wave mode with ~10mW being coupled into the PCF, as the pulse characteristic of the resulting 940 nm light will be governed by the pulse characteristics of the 1064nm pump light. The spectral output shown in Figure 1 shows all three peaks for performance assessment, however for imaging we filtered out the pump and idler light using a 1000 nm shortpass filter. Regarding FWM performance, we were able to generate signal peak powers of ~250 W at 940 nm with ~250 0W (peak) pump light at 1064nm coupled into the PCF. A further increase in pump power did not result in more signal light, but rather broadened the overall spectral peak at 940nm. This resulted in a peak power on the sample of 125 W at 940 nm. In order to generate higher signal peak powers in the future, we intend to use a larger core PCF fiber to lower unwanted non-linearities.

However, for the virtual H&E staining employed here, we noticed in agreement with [11] that a pump wavelength at 1030nm was already sufficient (see results and discussion). This wavelength lies within the Yb-gain profile and can thus be reached without employing FWM in the PCF. Therefore, the fiber MOPA presented in [14] was modified to generate 1030nm wavelengths as follows: The 1064nm seed diode was exchanged with a 1030 nm distributed feedback laser diode (Lumics LU1030M300). The laser line filters after the first and second stage were replaced with a home-built grating-based filter. They suffer from quite high losses of ~7dB due to losses from the grating and the fiber-to-fiber coupling. We plan to replace it with inline dichroic filter with significantly lower losses in the future. The double-clad (DC) stage was changed from a backward to a forward pumping architecture to minimize the travelling distance of the high power pulses in fiber and hence minimize unwanted stimulated Raman scattering (SRS) and stimulated Brioullin scattering (SBS). As we realized that reflections coming back from our microscope setup were damaging our laser, we added a free-space high-power isolator (Thorlabs IO-5-1064-HP) with fiber coupling. That last part also introduces quite high loses of about 6 dB. The currently used isolators in the stages are specified for 1064 nm and will be exchanged for models designed for 1030nm with lower losses and better isolation at 1030nm, which might render the last free space-isolator redundant. Further laser parameters used in this study were: 1 MHz repetition rate and 100 ps pulse width.

The TPM microscope setup has been previously described in [10] and is based on the one described in [15]. In brief, the light from the fiber MOPA is collimated with an 11 mm aspheric lens (Thorlabs A220TM-B) and deflected by an x- and y-galvanometric scanner (galvo, Thorlabs GVS002). The pivot points are relay imaged by a tube (Thorlabs AC508-075-B-ML) and a scan lens (combined Thorlabs AC508-300-B-ML and Thorlabs AC508-400-B-ML) into the back aperture of a 40x NA0.8 water immersion objective (Nikon N40XW-PF). There is a dichroic mirror with a 757 nm cutoff (AHF Analysentechnik HC BS 757) separating the epi-fluorescence and the excitation light before the objective. We used a non-descanned detection with a 40mm condenser lens (Thorlabs ACL5040) in front of a photomultiplier tube (PMT, Hamamatsu H12056-20). We mounted a filter wheel with three different fluorescence filters in front of the PMT: A 525nm bandpass with 39 nm FWHM (Thorlabs MF525-39), a 535 nm bandpass with 22 nm FWHM (Thorlabs MF535-22) and a 630 nm bandpass with 69 nm FWHM (Thorlabs MF630-69). The first two are used for collecting the AO fluorescence, whereas the primary one also transmits the second harmonic (SHG) signal of 1030nm, the third one is used for detecting the SR101 fluorescence. The signal from the PMT is digitized with a 12 bit 4 GS/s ADC-Card (Alazartech ATS9373). A home-built motorized xyz-stage beneath the microscope is used for creating large mosaicked virtual H&E images.

For comparison measurements, we use a commercial femtosecond two photon microscope (LaVision Biotec Tri M scope) equipped with Ti:Sa laser (Coherent Chameleon) and an OPO (Coherent APE Chameleon OPO). It is equipped with a 16x NA0.8 water immersion objective (Nikon N16XLWD-PF). The used fluorescence filters and PMT are the same type as in our fiber MOPA system.

We use chicken breast muscle, deer spleen and deer kidney as tissue samples. We adapted the staining protocol presented by Cahill et al. [11] and used 40 µg/ml AO and 40 µg/ml SR101 dissolved in 50% ethanol. As no temperature was specified, we decided to heat the solution to 40°C, which should enhance the chemical reaction but should not cause any changes to the tissue as it is within the range of body temperatures of these animals. The tissues were stained for 15 min and then washed for 1 min in 50% ethanol. Shorter staining times should be possible according to [11], but we wanted to ensure deep penetration of the stains into the tissue. The stained tissues were embedded in cavity slides before imaging.

The postprocessing of the data in our nsTPM-system is performed as follows: For each pixel, we record 70 digitizer samples corresponding to 12.5 ns, of which 10 digitizer samples capture the 2.5 ns before the excitation laser pulse and the remaining 60 digitizer samples (or 10 ns) capture the excited fluorescence signal. The first 10 digitizer samples can thus be used to determine a possible background or subtract a DC electronic offset caused by the ADC. For the 60 digitizer samples (10 ns), we first threshold samples by a single-photon voltage threshold such as to further reduce electronic noise originating from the digitizer card or the PMT. The 60 digitizer samples are then averaged to obtain the pixel intensity. Afterwards, the generated images are stitched together with a LabView program based on the known step size of the translation stage. Next, the mosaics are rescaled to 16 bit, whereas we applied a lower cut level in the AO image to enhance the visibility of the cell nuclei. The SR101 and AO images are acquired in sequential order due to having only a single PMT in our system at the moment, and are merged with the approach suggested by Giacomelli et al. [16].

## 3. RESULTS AND DISCUSSION

We start this section with an investigation of the effects of excitation wavelength on the signal levels and the specificity of the two stains and their emission wavelength. Then we present our virtual H&E images created with our fiber MOPA system and compare them with virtual H&E images created with the commercial femtosecond fiber laser system.

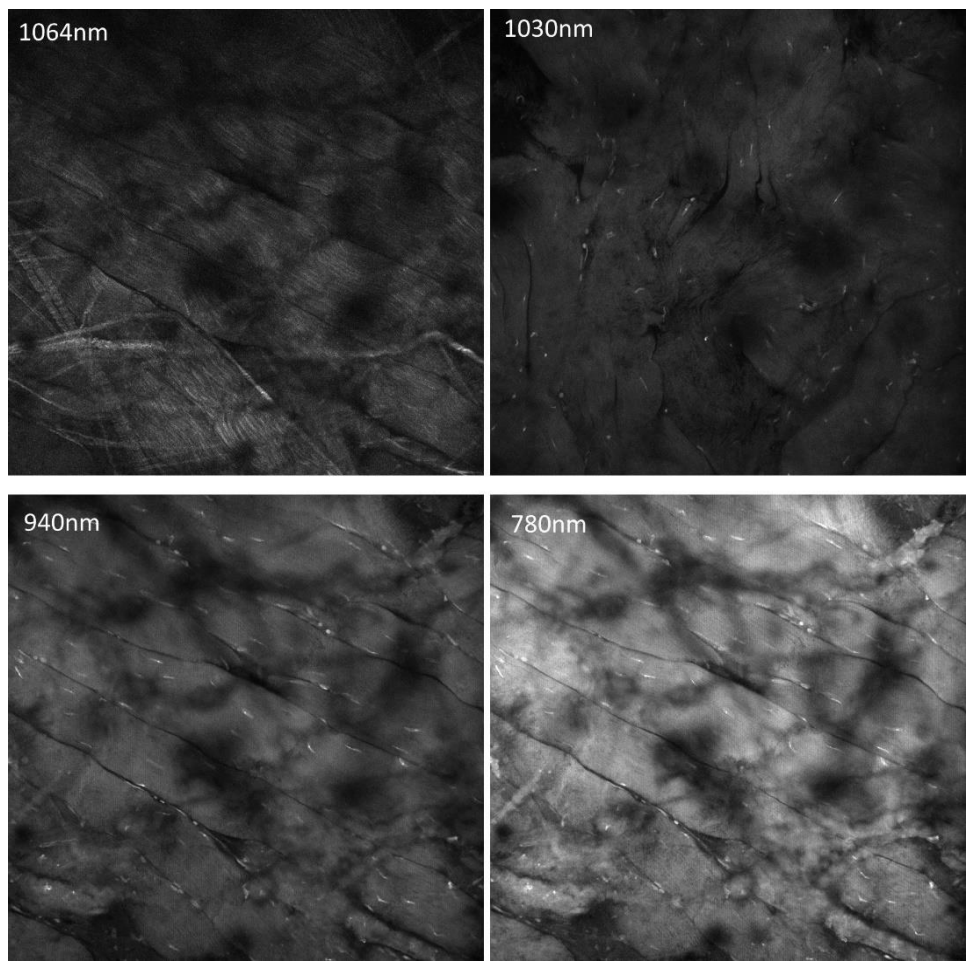

Figure 2: Chicken breast muscle stained with AO and SR101 imaged with a femtosecond TPM system at 50mW sample power and different wavelengths. The fluorescence light was filtered with the 525nm bandpass filter having 39nm FWHM. The image size is 1024x1024 pixels with a field of view of ~500x500µm.

As we reached quite low power levels at 940 nm, we decided to test different wavelengths in order to see whether it is necessary to increase our 940 nm power level or whether another wavelength is suitable as well for the excitation of AO and SR101. The critical stain is AO as it features a lower emission wavelength and therefore probably has a range of lower two photon excitation wavelengths. We imaged a sample of chicken breast muscle stained with AO and SR101 with the commercial TriMScope TPM system at 780 nm, 940 nm, 1030 nm and 1064 nm with ~50 mW cw-power on the sample and the 525 nm bandpass filter in front of the PMT. 780nm was chosen as it is a typical operation wavelength of a Ti:Sa laser. The 1064 nm were generated with the OPO of the system. The results are displayed in Figure 2. While at 1064 nm there is only SHG visible, but no nuclei, they can be seen at all other wavelengths with different intensities. An analysis of the histograms showed that the signal level at 940 nm is only a factor of about two higher than at 1030 nm, therefore only ~1.4 times more power would be required to reach the same amount of fluorescence considering the nonlinearity of TPM. Therefore we decided to use the more easily accessible and power-scalable 1030nm for virtual H&E imaging instead of 940 nm, leading to a more compact laser system.

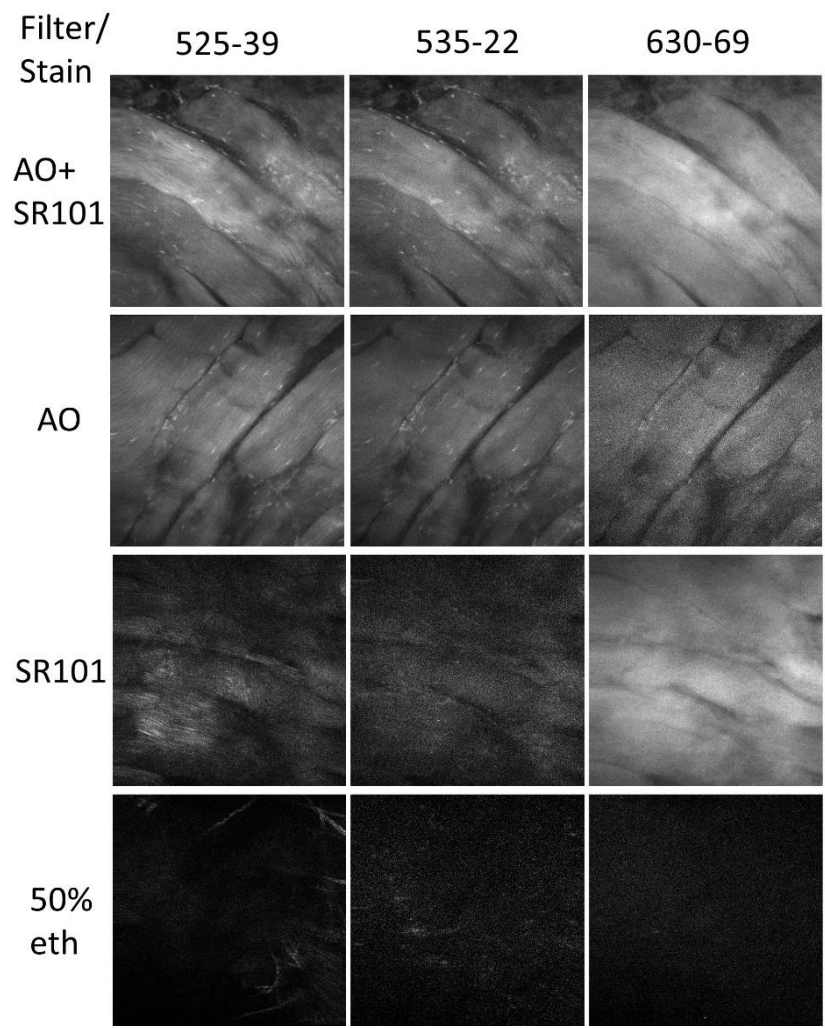

Figure 3: nsTPM-imaging of Chicken breast muscle. The tissue was stained using four different solutions: AO+SR101, AO only, SR only and a 50% ethanol solution. They were excited at 1030 nm. The fluorescence is filtered with three different bandpass filters, whereas the first number indicates the center wavelength and the second number the FWHM. Each image consists of 512x512 pixels and shows about 150 µm x 150 µm.

In order to understand the specificity of the AO and SR101, we stained four pieces of chicken breast muscle with each possible combination: AO and SR101, only AO, only SR101, and no stain. The concentration of 40 µg/ml was kept for each staining. The staining time was 3 min for each sample at room temperature as we did not optimize the staining procedure at this point. Each of the tissue samples was imaged with 1030 nm excitation light with the three filters mentioned above in front of the PMT. The results are displayed in Figure 3. The first column showing images acquired with the filter transmitting SHG exhibits fibrous structures, which could originate from collagen in the muscle. However, only signals from the cell nuclei are supposed to be in the AO image and therefore we chose the 535-22 filter for the AO images to block the SHG signal. Still, we see a significant amount of fluorescence from the structures around the nuclei (the elongated, bright structures). We assume that this is some unspecific staining of the AO. It can be suppressed in the final virtual H&E image by increasing the lower cutoff level (i.e. setting all signals below a certain threshold to zero), as the nuclei have a higher signal than their surroundings. Another observation is that some of the AO fluorescence is still transmitted through the 630-69 filter as seen in the sample only stained with AO. However, the SR101 fluorescence is significantly brighter and therefore no nuclei structures can be observed with the 630-69 filter in the dual stained sample. In future, maybe a filter with an increased lower cutoff should be chosen. The fluorescence of the SR101 is barely visible through the 535-22 filter chosen for the AO channel and should be removable by increasing the lower cutoff of the AO image similar to removing the unspecific AO background staining.

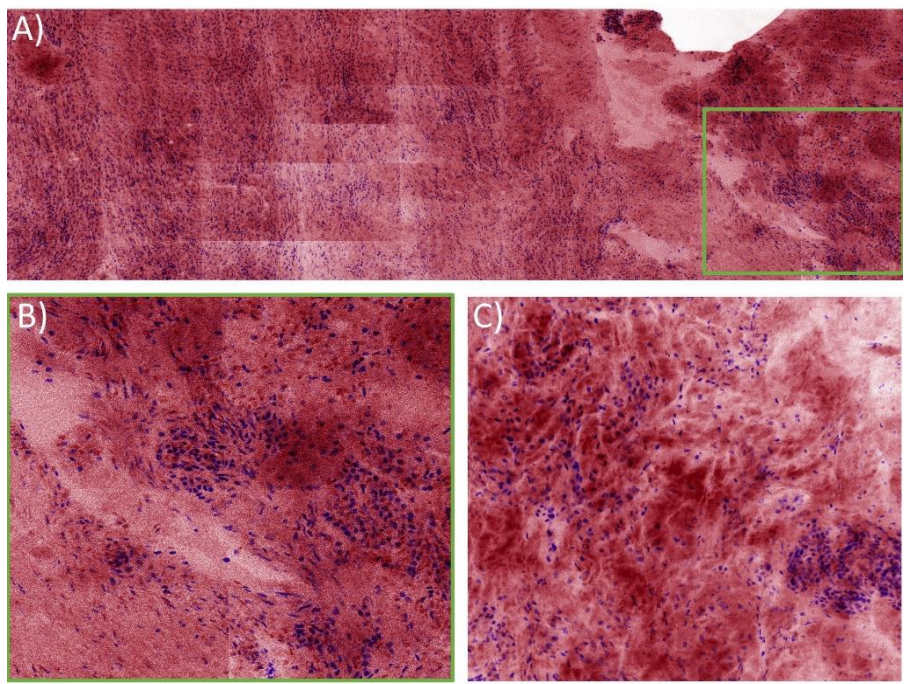

Figure 4. Virtual H&E mosaic of a deer kidney A) shows a ~2x0.8mm² large mosaic image consisting of 9x7 250x125µm² large tiles acquired with the fiber MOPA system. A pixel spacing of 500nm and 10% overlap between the tiles was chosen. B) is a ~500x500µm² zoom-in of the green marked area in A) showing individual cell nuclei. C) Shows a ~500x500µm² image of the same sample but at a different location acquired with the Ti:Sa laser system. Each image was acquired ~30µm below the sample surface.

A piece of deer kidney was prepared according to the methods section. We imaged it first with the commercial femtosecond system and then with our fiber MOPA. We adjusted the power levels and averaging such that we got comparable signal levels: As the fiber MOPA system is not optimized yet, we are limited to a peak power of 425 W at 100 ps and 1 MHz repetition rate at the sample. We applied 10x averaging to increase the signal level. The Ti:Sa laser has a repetition rate of 80 MHz and an estimated pulse duration of 200 fs at the sample. The femtosecond TPM system allows a pixel dwell time of 1.3µs corresponding to ~100 pulses per pixel. For comparison with the picosecond laser, the accumulated excitation can be considered, which can be calculated by the pulse peak power squared, times the pulse length times the number of pulses per pixel. Considering this figure of merit, the peak power of the femtosecond TPM system was chosen a factor of $\sqrt{5} \approx 2.23$ higher than the fiber MOPA system. To compensate for the 10x averaging with the nsTPM system, the previously computed peak power was increased by a factor of $\sqrt{10} \approx 3.33$ to 3 kW.

Figure 4 A) displays the virtual H&E mosaic of a deer kidney created with our fiber MOPA system (zoom-in in subfigure B)). There are still some artifacts of the stitching with our xyz-stage: The individual tiles have some vignetting leading to the discontinuities in brightness at some parts of the image. This is caused by minor clipping of the excitation beam at the back aperture of the objective lens while scanning due to a pivot plane mismatch between x- and y-galvo. In this case we aligned it closer to the x-galvo pivot point and chose an asymmetric scanning ratio. However, we would probably need to reduce the field-of-view further or chose a lower magnification of the objective lens. Moreover, some future work on the home-built xyz-stage positioning repeatability is needed, as can be seen by a few minor stitching errors at the mosaic borders.

Figure 4 C) shows a virtual H&E image of a different location of the same sample acquired with the Ti:Sa laser system. Comparing it to the image measured with the fiber MOPA system, the nuclei look quite similar, however, there seems to be more structure information visible in the SR101 counterstain for the Ti:Sa laser system. We cannot tell whether this is due to the different location being imaged, more signal from the SR101 stain or something else. A possibly higher signal could originate from the lower magnification of the objective lens used in the Ti:Sa laser system or from an actually higher peak power of the Ti:Sa laser system. The latter was only estimated by dividing the cw power by the duty cycle, whereas the pulse duration was not measured. A fair comparison would require a measurement of the pulse duration of the Ti:Sa laser and the use of the same objective lens, which however in this study was hindered by a different beam diameter in the commercial TPM microscope. We also notice for this sample and for the deer spleen, that the nuclei are faintly visible in the SR101 image. In contrast to our previous experiments shown in Figure 3, the SR101 signal does not outshine the AO signal passing through the 630-69 filter. We do not know whether this is due to the slightly different staining procedure or maybe was sample-specific. This is reproducible with both imaging systems.

## 4. CONCLUSION AND OUTLOOK

We present virtual H&E images acquired by a fiber MOPA based TPM system with 100 ps optical pulses and 1 MHz repetition rate. These images are compared to images acquired with a commercial TPM system with femtosecond pulses. Here, the counterstain images seem to provide more detail, but the reason is not clear. We plan to conduct a more thorough investigation by imaging the same location of the sample with both microscopes and also comparing the data to conventional H&E staining with either frozen sections and/or paraffin sections. Additionally, we could show the effects of certain excitation wavelengths reachable with a fiber MOPA system on the signal level. Furthermore, the specificity of the two stains and choice of suitable fluorescence filters was analyzed. In future, we will evaluate using a different filter for the SR101 channel.

Also, our fiber MOPA requires improvements to increase the effective pixel rate of currently 100 kHz (1 MHz repetition rate and 10x averaging per pixel). We plan to increase the available peak power at the sample, while decreasing the amount of averaging. We also intend to increase the repetition rate and decrease the pulse duration. Preliminary experiments showed that at least 30ps are possible with our fiber MOPA concept. Regarding our optical setup, we intend to use an objective lens with lower magnification for larger fields of view without vignetting. Currently, we also use unidirectional scanning with standard galvo scanners, which will be replaced with bidirectional scanning with resonant scanners in combination with the higher repetition rates. We will also implement a second PMT to enable parallel detection of the AO and SR101 fluorescence. Furthermore, we plan to analyze the already captured FLIM data to extract further information about the samples.

# ACKNOWLEDGEMENTS

This project was funded by the Hillenkamp Fellowship awarded by SPIE, the European Union (ERC CoG no. 646669), and the German Research Foundation (HU1006/6, EXC 306/2,), German Federal Ministry of Education and Research (BMBF no. 13GW0227B: "Neuro-OCT"),the European Union within Interreg Deutschland-Danmark from the European Regional Development Fund in the project CELLTOM. Ralf Brinkmann and Reginald Birngruber were Scientific and Translational Science Mentors for the Hillenkamp Fellowship of JPK.